# Time domain structures and dust in the solar vicinity: Parker Solar Probe observations

SHORT TITLE: TIME DOMAIN STRUCTURES AND DUST

**F.S. Mozer** (1,4), O.V. Agapitov (1), S.D. Bale (1,4,5,6), J.W. Bonnell (1), K. Goetz (2), K.A. Goodrich (1), R. Gore 9!,4) , P.R. Harvey (1), P.J. Kellogg (2), D. Malaspina (3), M. Pulupa (1), G. Schumm (1)

(1) University of California, Space Science Laboratory, Berkeley, CA, USA
(2) University of Minnesota, Minneapolis, MN, USA
(3) University of Colorado, Boulder, Laboratory for Atmospheric and Space Physics, Boulder, CO, USA
(4) University of California, Physics Department, Berkeley, CA 94720-7300, USA
(5) Imperial College London, The Blackett Laboratory, London, SW7 2AZ, UK
(6) Queen Mary University of London, School of Physics and Astronomy, London E1 K 4NS, UK

On April 5, 2019, while the Parker Solar Probe was at its 35 solar radius perihelion, the data set collected at 293 samples/sec contained more than 10,000 examples of spiky electric-field-like structures having durations less than 200 milliseconds and amplitudes greater than 10 mV/m. The vast majority of these events was caused by plasma turbulence. Defining dust events as those having similar, narrowly peaked, positive, single-ended signatures, resulted in finding 135 clear dust events, which, after correcting for the low detection efficiently, resulted in an estimate consistent with the 1000 dust events expected from other techniques. Defining time domain structures (TDS) as those having opposite polarity signals in the opposite antennas resulted in finding 238 clear TDS events which, after correcting for the detection efficiency, resulted in an estimated 500-1000 TDS events on this day. The TDS electric fields were bipolar, as expected for electron holes. Several events were found at times when the magnetic field was in the plane of the two measured components of the electric field such that the component of the electric field parallel to the magnetic field was measured. One example of significant parallel electric fields shows the negative potential that classified them as electron holes. Because the TDS observation rate was not uniform with time, it is likely that there were local regions below the spacecraft with field-aligned currents that generated the TDS.

Time domain structures (TDS) are several millisecond duration, intense, electron scale, electric field spikes having significant components parallel to the local magnetic field [Mozer et al., 2015; Hutchinson, 2017]. At least five different types of TDS exist including electrostatic and electromagnetic double layers, electrostatic and electromagnetic electron holes, and nonlinear whistlers [Mozer et al, 2015; Drake et al, 2015; Agapitov et al, 2018]. Electron hole TDS were first studied in numerical models of the instability of two electron beams [Roberts and Berk, 1967; Morse and Nielson, 1969a, 1969b]. Saeki et al,[1979] first observed their generation in the lab in a Q-machine. Generation by spontaneous reconnection in a toroidal device was studied [Fox et al., 2008], and laboratory observations of electron holes in a plasma device have been presented [Lefebvre et al., 2011]. Double layer-type TDS were first discussed in connection with magnetospheric physics and astrophysics by the Stockholm group under Hannes Alfvén [Alfven and Carlqvist, 1978, and references therein; Raadu, 1989], and they were first observed in the magnetosphere along auroral zone magnetic field lines by the S3-3 satellite [Mozer et al., 1977; Temerin et al., 1982]. They were more thoroughly studied on the FAST mission [Ergun et al., 1998a]. They have been seen in the magnetospheric tail [Matsumoto et al., 1994; Franz et al., 1998; Streed et al., 2001], the plasma sheet [Ergun et al., 2009; Deng et al., 2010], the plasma sheet boundary layer [Lakhina et al., 2010], at shocks [Bale et al., 1998; Cattell et al., 2003], at magnetic field reconnection sites [Cattell et al., 2002; Mozer and Pritchett, 2009; Khotyaintsev et al., 2010; Li et al., 2014], in the solar wind [Bale et al., 1996; Malaspina et al., 2013a; Williams et al., 2005], and at Saturn [Williams et al., 2006]. They have been invoked to explain particle heating, scattering and acceleration [Ergun et al, 1998b; Bale et al, 2002; Cranmer et al, 2003; Mozer et al, 2016; Vasko et al, 2017,2018].

Dust impacts have been observed on many spacecraft in the solar wind since they were first found in Saturn's rings [Aubier et al, 1983; Gurnett et al, 1983; Meyer-Vernet et al, 2009, 2014, 2017; Malaspina et al, 2013b; Zaslavsky, 2015; Vaverka et al, 2018]. Szalay et al [2019] and Page et al [2019] describe properties of dust striking the Parker Solar Probes. Their dust data was obtained from microsecond resolution, infrequently recorded bursts of fields data. Another view of the same phenomenon is obtained by analyses, reported here, of lower time resolution (293 samples/second) full time coverage of the spikes observed in the electric field detector on the Parker Solar Probe. An additional result of this analysis is the discovery of abundant time domain structures and the determination of their properties. The electric field detector is described elsewhere [Bale et al, 2016].

Figure 1 presents >10 Hz filtered data collected by the Fields DFB waveform receiver at 293 samples/second on April 5, 2019, at the 35 solar radius perihelion of the spacecraft orbit. The data are in the spacecraft coordinate system which has X perpendicular to the Sun-spacecraft line, in the ecliptic plane, and pointing in the direction of solar rotation (against the ram direction), Y perpendicular to the ecliptic plane, pointing southward, and Z pointing sunward. The Z-component of the electric field was not measured. The four single ended potentials V1, V2, V3, and V4, whose differences (V1-V2) and (V3-V4) determine the X and Y components of the electric field, are in a plane parallel to the satellite heat shield. In Figure 1 there are 9800, 10-20 mV/m amplitude, electric field spikes with durations <200 milliseconds, 750 20-30 mV/m spikes, and 760 >30 mV/m spikes. By far, the largest fraction of these events resulted from differences in sensor floating potentials driven by density fluctuations of the type illustrated in Figure 2. Panels 2a and 2b of Figure 2 give the apparent electric field measured due to potential differences between the single-ended antenna measurements V1 (not shown because it was measured at a too-low frequency) and V2 (panel 2c), as well as V3 (panel 2d) and V4 (panel 2e). Although the single-ended potentials all have the same shape, the ~20% differences in their 100 millivolt amplitudes produces the apparent 20 mV/m electric field seen in panels 2a and 2b. The task in the following data analysis is to distinguish between these turbulent electric fields, those produced by dust, and those produced by TDS.

Figure 3 illustrates a dust impact seen simultaneously in the sub-microsecond resolution channels (panels 3a through 3d), and the 293 samples/second channels (panels 3e through 3j). As seen in this figure, a dust hit produces a <0.1 msec electric field spike in the single ended potentials V1, V2, V3, and V4, when these voltages are measured with sufficiently high time resolution (panels 3a, 3b, 3c, and 3d). This spike results from the rapid release of spacecraft ions by the impact [Meyer-Vernet et al, 2009]. Because this release causes the potential of the spacecraft to change, V1 through V4 experience similar changes because their measured quantity is the potential of the antenna minus the potential of the spacecraft. The system recovers from this dust hit in the time required to charge the spacecraft capacitance to its previous condition. This charging of the spacecraft capacitance depends on the spacecraft shape and the plasma thermal current, but it is typically the order of one millisecond. This signal is processed by electronics that produce an overshoot for two reasons. First, the data is AC coupled, which results in the average signal being zero. So the spike must be followed by a signal of the opposite sign and identical area. The shape and duration of this overshoot depends on the electronics. Second, if the spike is sufficiently large, it saturates the electronics, which then take longer to recover, depending again on the electronic design. Panels 3g through 3j are the high rate data of panels 3a through 3d after passing through a 100 Hz low pass filter associated with the lower data rate. This filter both attenuates and spreads the input signal in time due to the fact that the low rate channel cannot see components of the input signal faster than ~10 msec. For this reason, a millisecond dust pulse generally appears in the 293 samples/sec, single-ended voltage channels as a single point peak in a >10 msec signal with a following overshoot whose duration depends on the amplitude of the signal and the electronic design.

The single-ended potentials in panels 3g through 3j are similar in shape but enough different in amplitude to produce the apparent 10 mV/m electric field in panels 3e and 3f. To understand how general the data of Figure 3 are, 104 dust spike plots of V2 through V4 (V1 was not measured), measured at 293 Hz at times when known dust pulses were simultaneously observed in the sub-microsecond resolution data, were studied. They show that, in 72% of the cases, the single-ended signals were like those illustrated in panels 3h, 3i and 3j, lasting ~100 milliseconds although the dust signature at high time resolution (panels 3a through 3d) lasted ~one millisecond. While there are dust impacts that do not fit this description, this empirical result is used to define dust events in the 293 Hz DFB channel as those for which all the positive, sharply peaked and similarly shaped but different amplitude single-ended potentials with <200 milliseconds durations [such as those of panels 3h through 3j are observed. This resulted in a detection efficiency of 72%. Previous observations of dust impacts on other spacecraft differ from this definition because such data were measured at different data rates, involved other than the several single-ended potentials, did not involve current-biased antennas, and involved electronics having different saturation properties from those on the Parker Solar Probe.

Figure 4 presents another dust event observed at 293 samples/second. As expected from the above discussion, the three single-ended voltages are positive, peaked at a single point, and somewhat different in amplitude (panels 4c, 4d, and 4e). In this case, the apparent electric field of panel 4b was 1000 mV/m, which is difficult to understand unless the input pulse saturated the low frequency electronics. As seen in the two dust examples of Figures 3 and 4, the apparent electric fields due to dust can be small (~10 mV/m) or, occasionally larger than 1000 mV/m. Because the dust duration is defined as the time interval during which the 293 Hz signal exceeds a fixed voltage threshold, the observed durations depend mainly on the amplitude and duration of the overshoot that is discussed above. For example, the differential signal in Figure 3e and 3f is above a 5 mV/m threshold for about 15 msec while the signal in Figure 4a and 4b is above the same threshold for more than 100 msec.

Figure 5 presents an example of an electric field signal produced by a TDS. Panels 5a and 5b of this figure give the electric field components in the X-Y plane, and the remaining panels give the single-ended voltages V2, V3, and V4. What is uniquely different from dust pulses in this example is that the three single-ended voltages are not similar in form. Instead, V3 and V4 have opposite polarity components, which means that there is a real electric field across the spacecraft because the antenna on one side of the spacecraft has an increasing (decreasing) voltage at the same time that the opposite antenna has a decreasing (increasing) voltage. Furthermore, the electric field signal is bipolar, which is the known structure of most of the previously observed TDS in the solar wind and elsewhere throughout the heliosphere. The TDS duration is the order of several Debye lengths divided by the electron thermal velocity, which is ~50 msec for the plasma parameters at the observation time. The similarity of the dust and the TDS durations is a coincidence associated with the plasma parameters for the TDS and the electronic design and data rate of the dust.

Using the requirement that dust produces similar and narrowly peaked single-ended potentials on the three antenna while TDS have opposite polarity V3 and V4 values, the 1500 examples of >20 mV/m electric fields observed on the day of interest were examined to determine those due to dust and those due to TDS. The result of this search is shown in Figure 6, which plots the amplitude and width of each event. As expected, dust produced both the largest and smallest amplitude events and the longest and shortest duration events because the dust signature depended on the overshoot and saturation properties of the incident signal and the electronics, as discussed earlier. 238 TDS and 135 dust strikes were found in this data set.

To further characterize the TDS, the bipolar nature of each event is plotted in Figure 7 as the ratio of the amplitude of the first peak to the amplitude of the opposite polarity peak, with the sign of the ratio being the sign of the first peak. Most bipolar structures had roughly equal amplitudes of their positive and negative maxima and the negative polarity maximum appeared before the positive polarity maximum in the majority of cases. This latter result may be an artifact of the selection of TDS using only voltages V3 and V4, or it may contain information on the TDS propagation direction.

The TDS data were searched for events that occurred when $B_Z$ was close to zero because, for such events, the magnetic field was in the X-Y plane (as was the measured electric field) and the component of the electric field parallel to the magnetic field could be determined. Figures 8a and 8b give one example of the perpendicular and parallel electric fields measured when the magnetic field was within 3 degrees of the X-Y plane. Panel 8c gives V3 and V4, which have opposite polarity components, as is required for a TDS. That V3 and V4 are dissimilar in magnitude is the result of the potential of the spacecraft changing during the event to add to one of the signals and to subtract from the other. Assuming that the TDS was moving in the $-B_Z$ direction, away from the sun, this is an example of a negative potential structure, which makes it an electron hole.

Figure 9 gives the TDS observation rate versus time on the day of interest, during which the spacecraft distance from the sun varied by less than one solar radius. This non-uniform distribution suggests that TDS events appeared in spatially or temporally confined regions. It also shows that this distribution was not produced by dust impacts or turbulence because they would be uniform in time.

DISCUSSION

Time domain structures and dust events are identified for one day at the 35 solar radius perihelion of the Parker Solar Probe, by examining the single-ended voltages on three electric field antennas in the X-Y plane when the amplitude of the electric field exceeded 20 mV/m. The criterion for identifying a dust impact was that the three antennas have similar positive, narrow, voltage peaks and the criterion for a TDS was that V3 and V4 had opposite polarity components. 238 time domain structures were found in this way. The detection efficiency for TDS was less than 50% both because turbulence (plasma density perturbations) can be large enough to hide the anti-correlation and because only V3 and V4 were used for identification. Considering these effects, it is estimated that between 500 and 1000 TDS were present on this day. This estimate is supported by observation of a similar number of bipolar electric field events to those in the 238 identified events although they did not have opposite polarity V3 and V4 signals. Such events may have had opposite V1 and V2 signals.

For a 73% detection efficiency of >20 mV/m dust events, the 135 identified dust events correspond to a probable 185 dust events in the day of data. This estimate is low because most of the dust impacts had amplitudes <20 mV/m (Figure 3) and were lost in the huge number of turbulence events. The number of dust events estimated in other ways [Szalay et al, 2019; Page et al, 2019] is suggested to be about 1000. Thus, because most dust events were too low in amplitude to be detected by the detection technique in this paper, as illustrated in Figure 2, the observation of 185 events with amplitudes >20 mV/m seems reasonable.

A few time domain structures were observed at times when $B_Z$ was small, such that the magnetic field was in the plane of the electric field measurement and the component of the electric field parallel to the magnetic field was measured. Under the assumption that the TDS was moving away from the Sun, such structures were electron holes, in agreement with TDS properties in the magnetosphere [Mozer et al., 2015, 2016; Vasko et al., 2017,2018] and from numerical simulations [Drake et al., 2015]

Because the TDS appeared in spatially confined regions and because electron holes are likely associated wih field-aligned currents [Mozer et al, 2015; Hutchison, 2017], this data suggests the presence of regions containing field-aligned currents at lower altitudes. While electron holes can significantly heat and accelerate electrons [Ergun et al, 1998b; Bale et al, 2002; Cranmer et al, 2003; Mozer et al, 2016; Vasko et al, 2017,2018], they may not occur in sufficient numbers at 35 solar radii to cause a local effect. However, closer to the Sun, in regions that will be examined in later orbits, their numbers might increase to levels of interest for the electron dynamics.

ACKNOWLEDGEMENTS

This work was supported by NASA contract NNN06AA01C. The authors acknowledge the extraordinary contributions of the Parker Solar Probe spacecraft engineering team at the Applied Physics Laboratory at Johns Hopkins University. SDB acknowledges the support of the Leverhulme Trust Visiting Professor program.

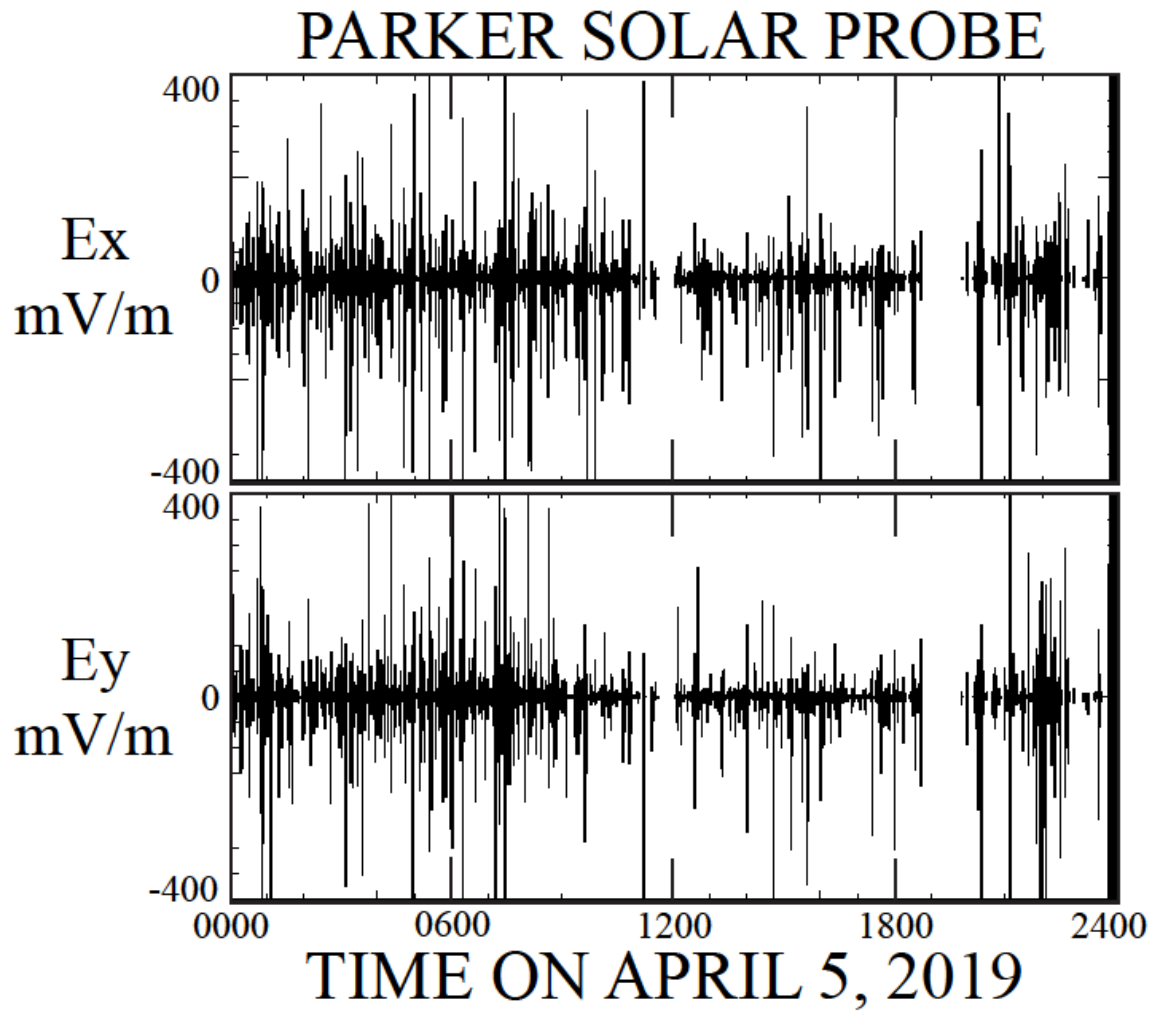

Figure 1. One day of high pass (>10 Hz) electric field data obtained at the 35 solar radius perihelion of the Parker Solar Probe orbit. The distinguishing feature in the data is the many short duration spikes in the electric field caused by turbulence, time domain structures and dust impacts on the spacecraft body.

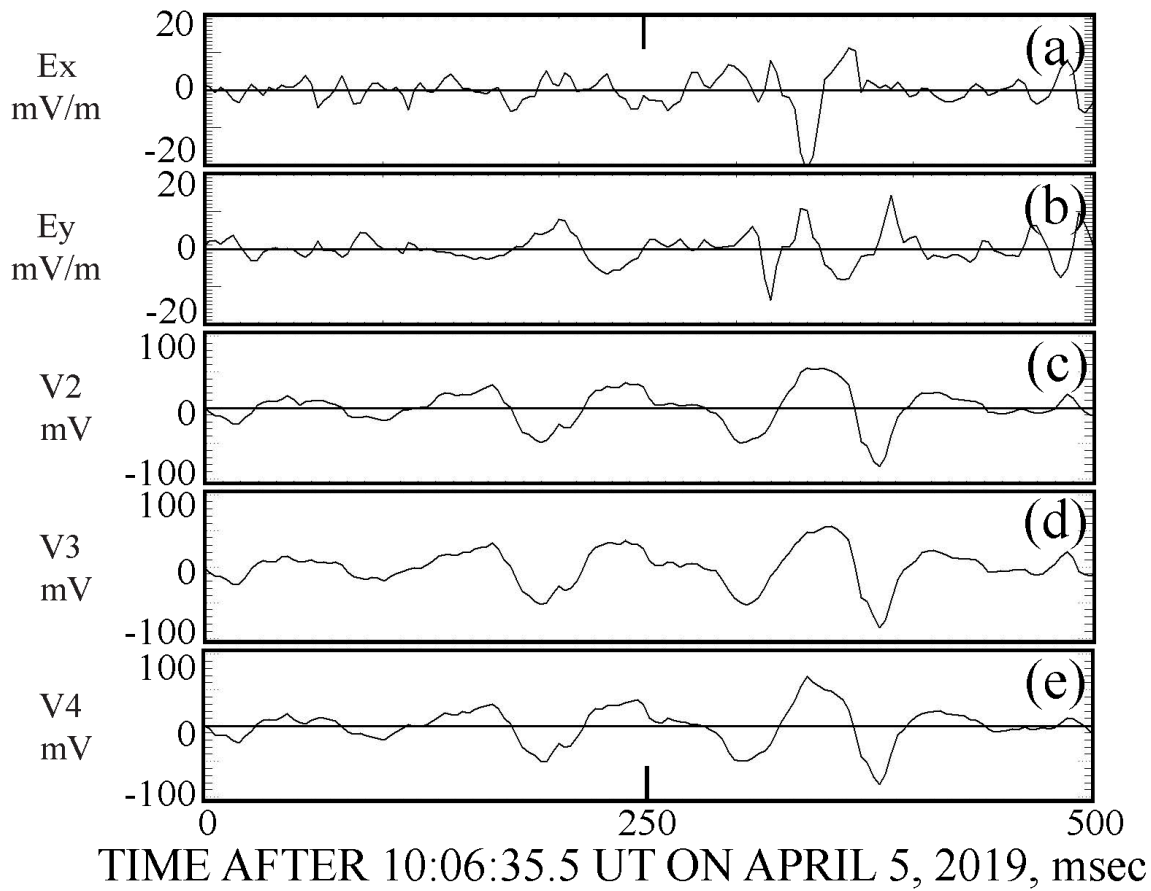

Figure 2. Panels 2a and 2b give the apparent electric field measured by the potential differences between the single-ended measurements V1 (not measured) and V2 (panel 2c), as well as V3 (panel 2d) and V4 (panel 2e). Although the single-ended potentials have the same shape, their ~20% differences in their 100 millivolt amplitudes produces the apparent 20 mV/m electric field seen in panels (a) and (b).

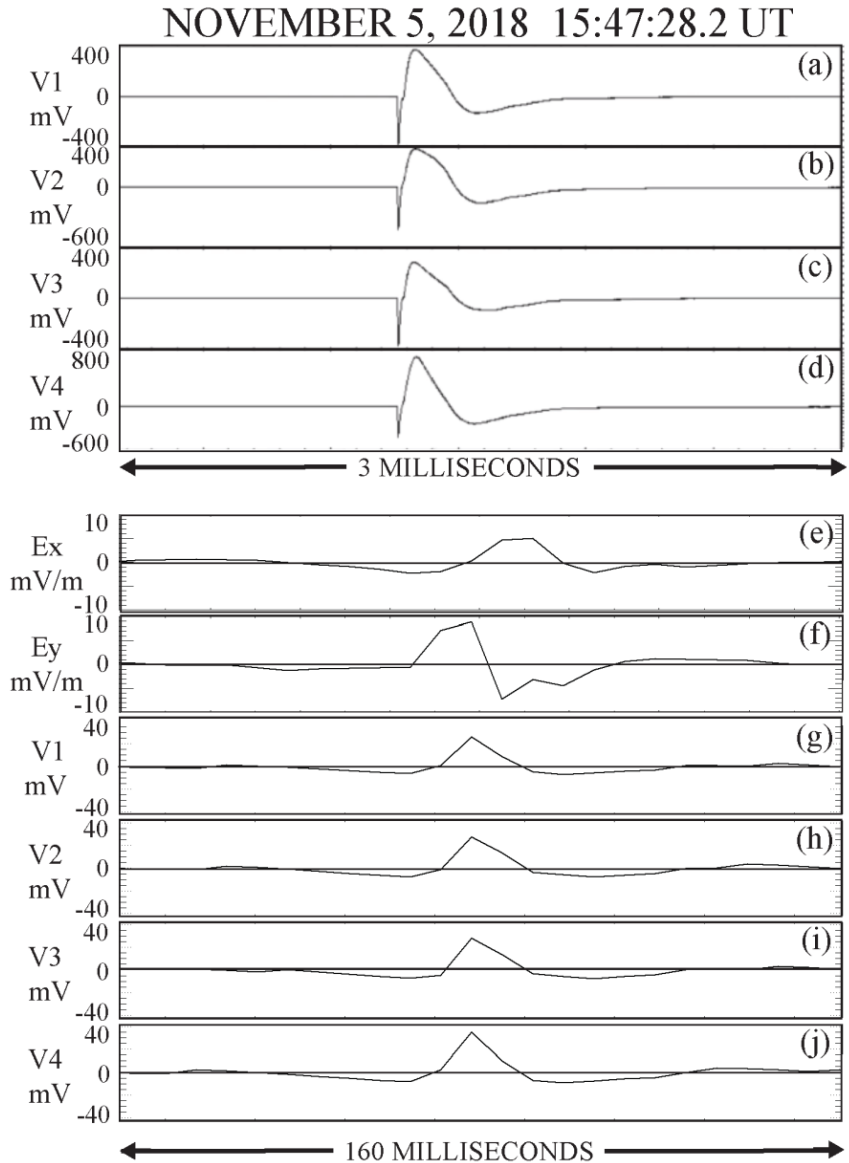

Figure 3. Panels 3a through 3d give three milliseconds of high time resolution single-ended voltages measured on the four individual antennas during a dust impact. Significantly, the signals on all of the antennas are essentially identical. Panels 3e and 3f3 give 100 milliseconds of the <10 mV/m electric field signals measured at 293 Hz for the dust event and panels 3g through 3j give the four antenna signals, which are also essentially identical.

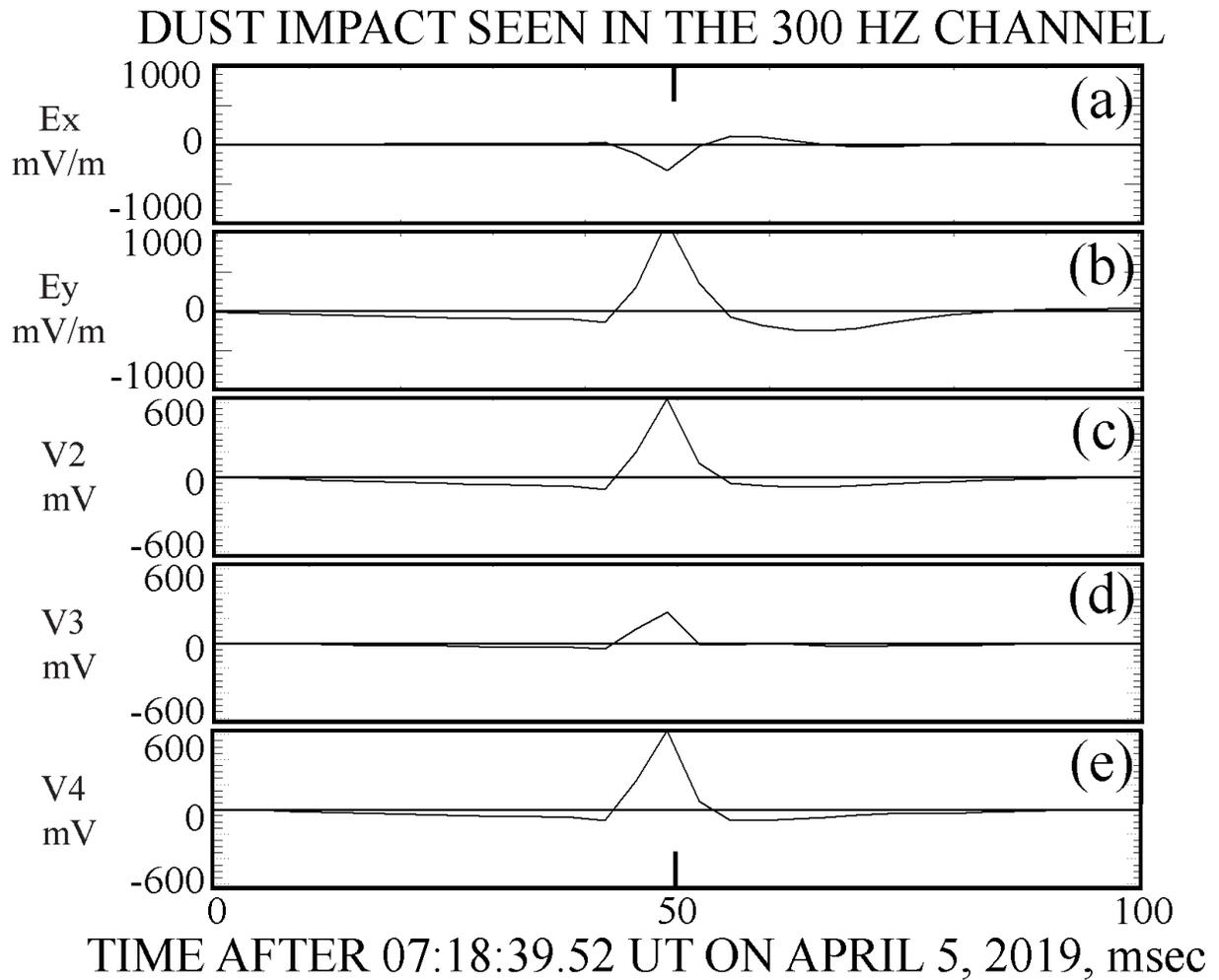

Figure 4. A huge (>1000 mV/m) electric field signature of dust seen at 293 Hz along with three of the four individual antenna voltages. This signal presumably resulted from saturation of the measuring electronics.

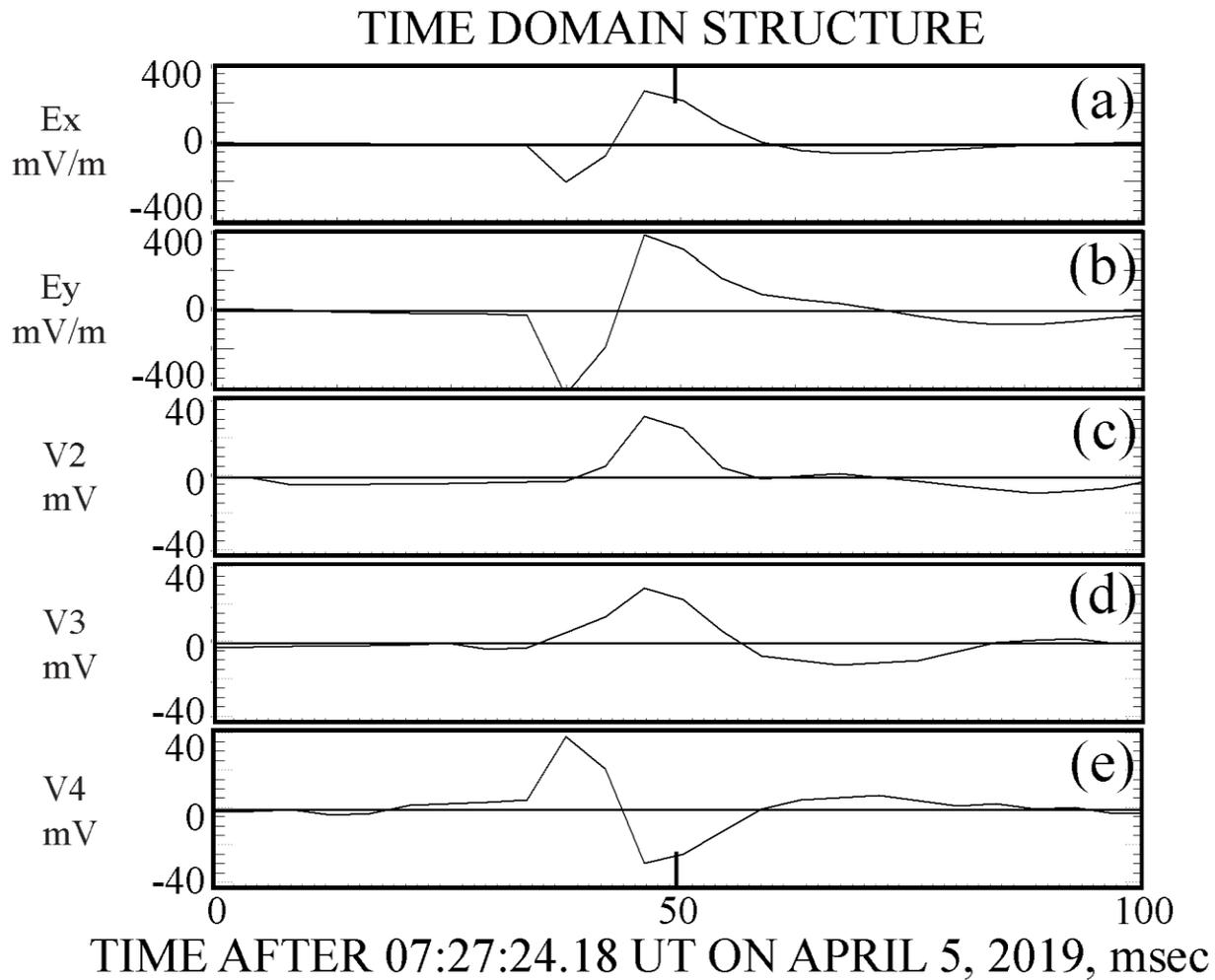

Figure 5. The electric field (panels 5a and 5b) and the individual antenna signals (panels 5c through 5e) for a time domain event, which is distinguished from a dust event because the signals from the opposite antennas, V3 and V4, have opposite polarity components..

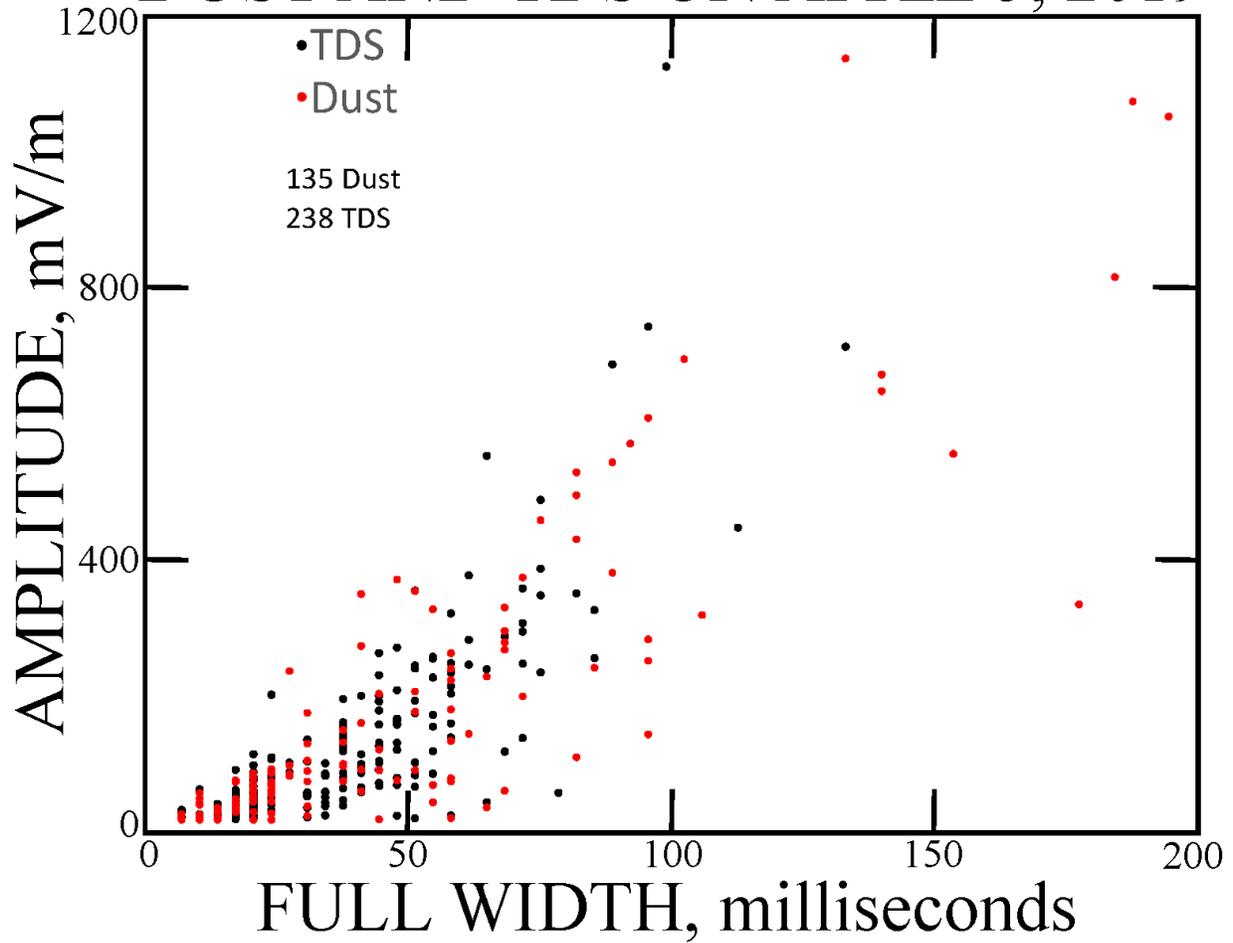

Figure 6. The distribution of amplitudes and widths of dust and time domain structures.

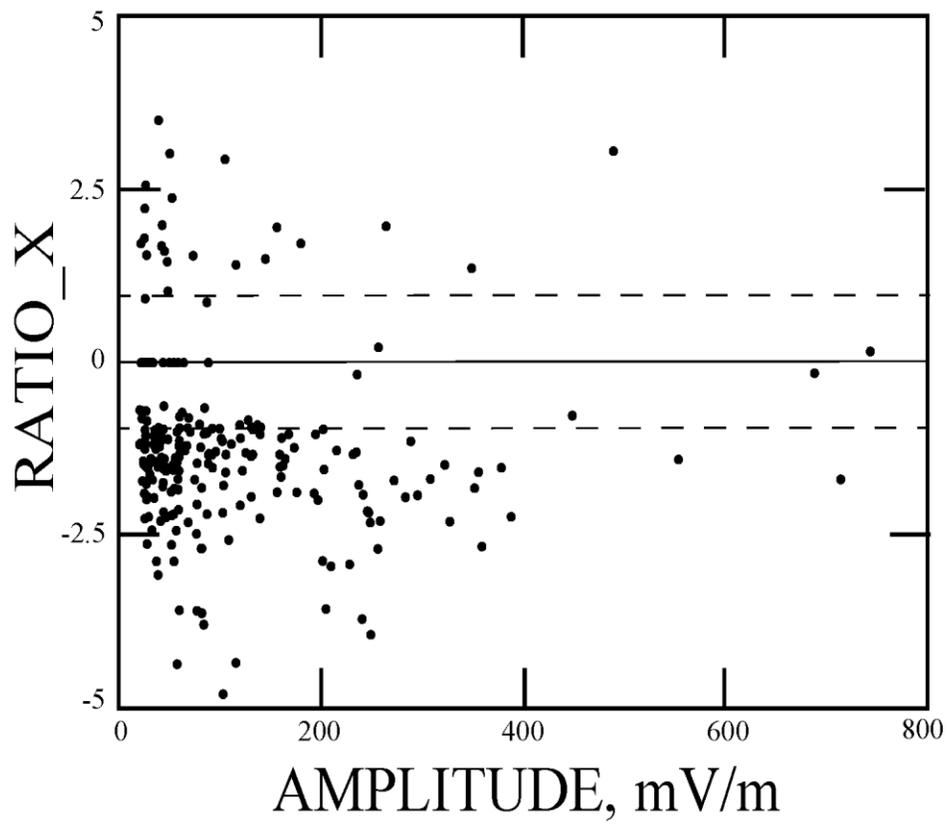

Figure 7. The ratio of the first peak of a bipolar TDS signal to the second peak, with the sign of the ratio being the sign of the first peak.

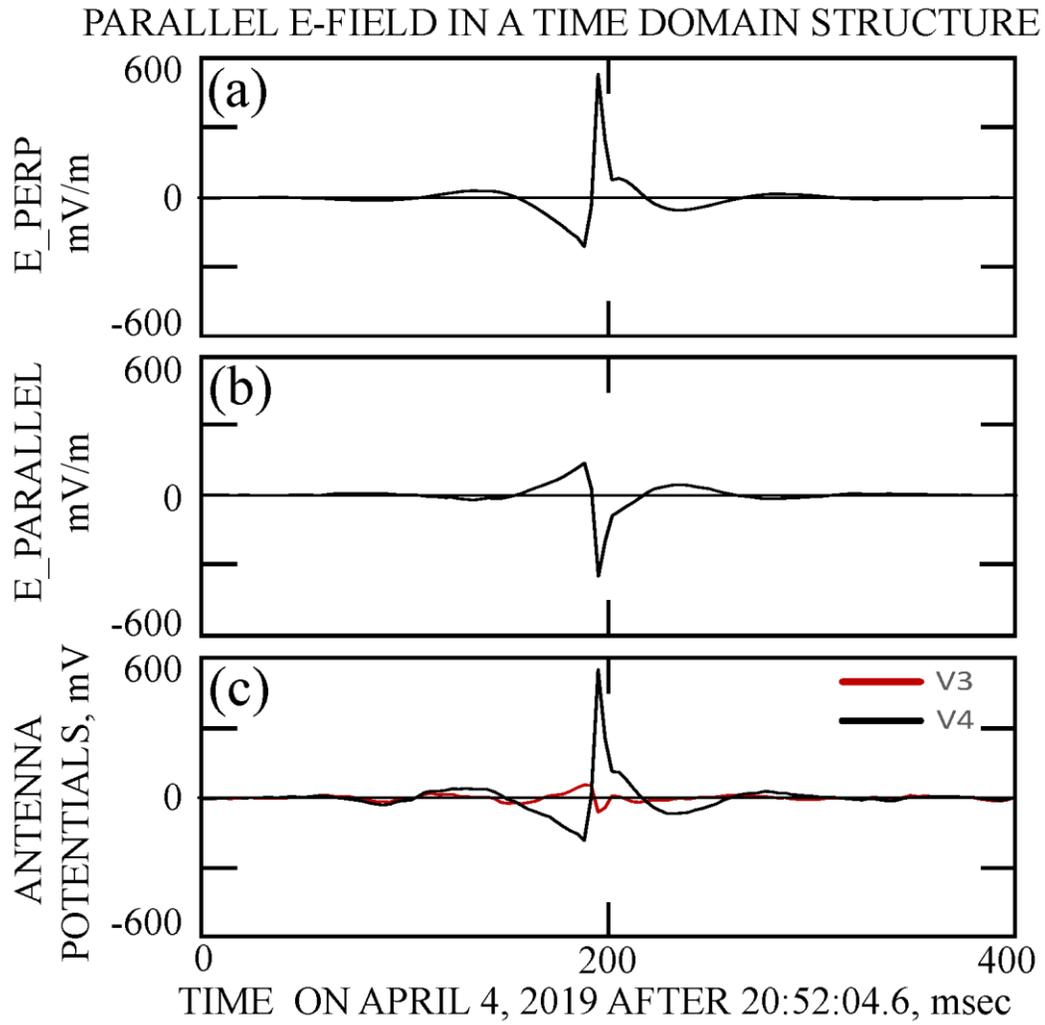

Figure 8. The electric field components perpendicular and parallel to the local magnetic field in a time domain structure.

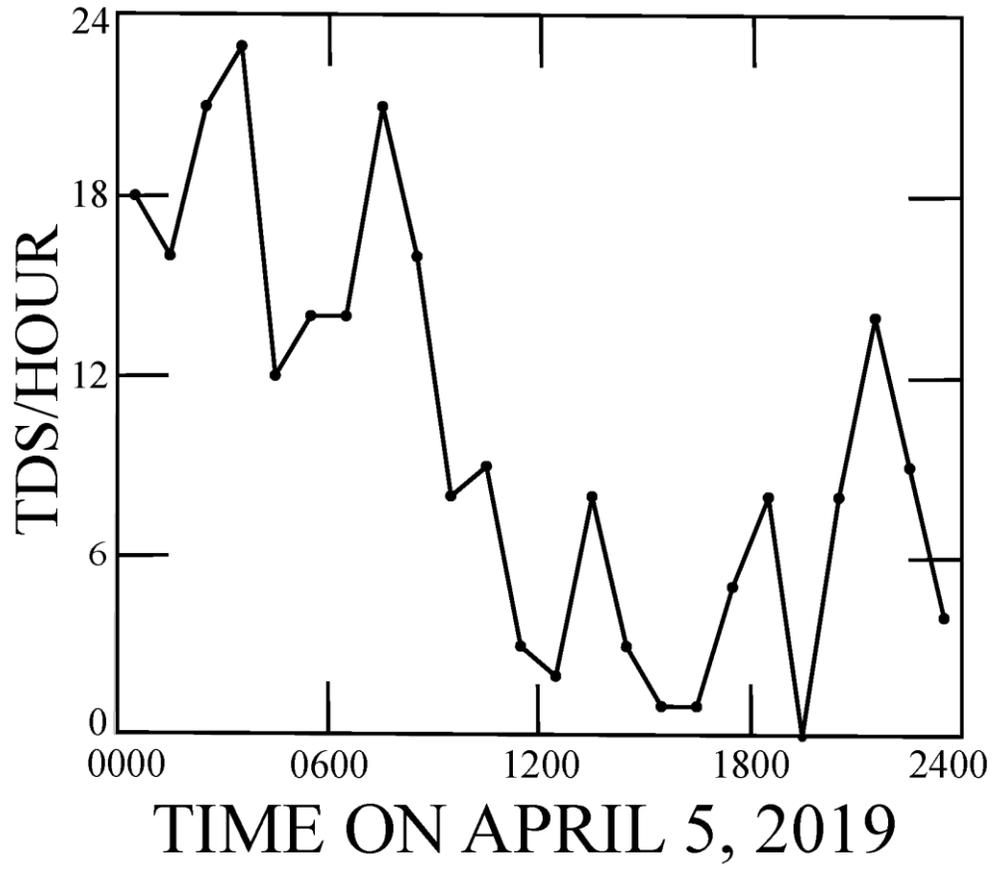

Figure 9.   The TDS rate versus time.